# Influence of nanoparticulates and microgrooves on the secondary electron yield and electrical resistance of laser-treated copper surfaces

P. Krkotić,[*] T. Madarász,[†] C. Serafim,[‡] H. Neupert, A. T. Perez-Fontenla,
M. Himmerlich, and S. Calatroni[§]

CERN, 1211 Meyrin, Switzerland

S. Wackerow and A. Abdolvand

*School of Science and Engineering, University of Dundee, DD1 4HN Dundee, Scotland*

Laser surface structuring has proven to be an effective technique for achieving a copper surface with secondary electron yield (SEY) values close to or below unity. However, the attributes that minimize SEY, such as moderately deep grooves and redeposited nanoparticles, may lead to undesirable consequences, including increased radio frequency surface resistance. This investigation systematically examined data about different cleaning procedures designed to eliminate redeposited adsorbed particulates. Various analysis techniques were used iteratively after each consecutive cleaning step, providing insights into the evolving surface characteristics. The collected experimental results identified distinct impacts of microgrooves, groove orientation, and associated particulates on secondary electron yield and surface resistance. Exposing the crests while retaining high particulate coverage in the grooves leads to reduced SEY values and surface resistance, suggesting that the tips of the grooves exert a more significant influence on surface current density than the groove depth. At the same time, nanoparticles in the grooves have a more significant impact on SEY values than the exposed tips at the surface.

DOI: 10.1103/PhysRevAccelBeams.27.113101

## I. INTRODUCTION

The electron multipacting and, consequently, the electron cloud effect are known phenomena that can occur in particle accelerators. Photoemission from synchrotron radiation, ionization of residual gas, and perhaps uncontrolled loss of stray beam particles all contribute to generating primary electrons which may impinge on the beam vacuum chamber leading to the emission of secondary electrons. These can be accelerated in the electromagnetic fields generated by the particle beams and multiplied in avalanche when specific resonance conditions occur. Such a cloud of electrons can attract and accumulate more charged particles and thus can have several detrimental consequences on accelerator performance, including increased beam emittance, instabilities, and beam loss.

The effects of multipacting and electron clouds are crucial for upcoming circular colliders aiming for proton beams with higher intensity and energy [1–9].

The electron multipacting process occurs in general in the presence of intense radio frequency (rf) electric fields, typically found in accelerator and also in satellite components like waveguides, coaxial structures, filters, or resonant cavities [10]. When the rf field reaches critical strength, it has the potential to induce a resonant acceleration and deceleration of electrons within the structure, which may lead to the generation of secondary electrons causing increased power losses, rf breakdown, and degradation of performance. Thus, in both accelerators and satellites, managing the electron multipacting and cloud effect is crucial to maintain the desired performance and reliability of the systems.

Various strategies have been investigated and implemented to address these adverse effects of electron multipacting. One approach involves shaping the geometry of rf structures to create unique configurations that mitigate multipacting effects [11,12]. The other common mitigation technique considers minimizing the vacuum components' SEY through careful material selection, special coatings, or other physical or chemical surface treatments [13]. The SEY value is defined as the number of secondary electrons leaving a surface upon the incoming primary electrons number and when this ratio exceeds unity, electrons have the potential to multiply, whereas a surface with a SEY below unity functions as an electron absorber.

---

[*]Contact author: patrick.krkotic@cern.ch
[†]Also at Budapest University of Technology and Economics, 1111 Budapest, Hungary.
[‡]Also at University of Helsinki, 00100 Helsinki, Finland.
[§]Contact author: sergio.calatroni@cern.ch

Several studies have demonstrated that laser surface treatment on copper can significantly decrease the SEY, depending on the various laser and treatment properties [14–18]. Such laser treatment is the controlled and precise modification of a material's surface using laser beams to generate specific patterns, textures, or structures. Laser ablation surface engineering [14], laser-induced periodic surface structures [15], and laser-engineered surface structures (LESS) [16], each of which uses a different set of laser-treatment parameters, are suitable treatment techniques to reduce the SEY value of copper.

Besides a low SEY value, maintaining an appropriately low surface impedance in both accelerator and satellite components is essential for optimizing performance and minimizing power dissipation. In accelerators, the surface impedance influences the electromagnetic interaction between the circulating beam and the surrounding vacuum components. Therefore, a low surface impedance minimizes impedance-driven beam instabilities. More in general, a lower surface impedance of vacuum components, such as waveguides or resonators, permits the maintenance of signal integrity. As a consequence, there is a keen interest in methods that can reduce SEY levels while preserving favorable rf characteristics.

Laser structuring might raise the surface resistance of the treated material [19,20]; though, independent tests on these laser-structured surfaces at mega- and gigahertz frequencies at room and cryogenic temperatures have revealed that the surface resistance varies greatly depending on the relative orientation of the laser scan lines with respect to the rf surface currents, on the depth of the groves produced, and, more generally, on the overall topography [15,19,21,22]. Particularly, LESS, due to its production method, results in a highly complex surface topology achieved by repeatedly scanning a laser beam in a linear pattern [16]. The individual contributions of the formed microgrooves and the attached particles to the change of the SEY and the surface resistance of laser-treated materials still need to be understood in detail [23]. Therefore, in this study, the step-by-step removal of nanostructures on a laser-treated sample aims to explore these relations.

This investigation systematically examines various cleaning procedures designed to remove the surface particulates stepwise, enabling fine adjustments to the topography of laser-treated materials in order to measure the influence on the SEY and on the surface resistance. The results of these measurements are supported through a combination of optical microscopy and topography mapping for geometrical aspects, scanning electron microscope (SEM) analysis, and chemical analysis via x-ray photoelectron spectroscopy (XPS), iteratively performed after each cleaning step.

## II. SAMPLE PREPARATION

### A. Laser structuring

Laser treatment was applied to two sets of bulk oxygen free electronic (OFE) copper discs with two distinct dimensions for the various experimental configurations: 5 cm diameter for SEY, XPS, and SEM analysis and 12 cm for surface resistance measurements and optical inspection. As visible in the magnified optical micrographs in Fig. 1, in the first set, denoted as LESS-I, the grooves were predominantly engraved in the radial direction, forming triangular sectors. In contrast, in the second set, referred to as LESS-II, lines were etched in a continuous spiral emanating from the center of the disc. Each disc of the same set has been treated with the same laser parameters independent of the disc dimension. The laser surface structuring process was performed using a 10 ps pulsed laser with a wavelength of 532 nm, linear polarization, and a repetition rate of 200 kHz. The structures were made using patterns with a line distance of roughly 45 μm at 15 mm/s laser spot scanning speed. In order to focus the laser beam onto the surface, a lens system was employed in conjunction with a nitrogen nozzle. By operating the laser at its maximum power, a Gaussian intensity profile with a standard deviation of $4\sigma$ and a spot diameter of 52 μm, an average fluence of approximately 0.9 J/cm$^2$ or an average equivalent power of 4 W was achieved. In order to minimize possible surface oxidation, a controlled flow of nitrogen was directed toward the light-matter interaction region in a laminar manner. The detailed procedure can be found in [16]. These two laser treatment configurations are selected based on the surface currents induced during the experimental examination of the surface resistance explained in detail in Sec. III E.

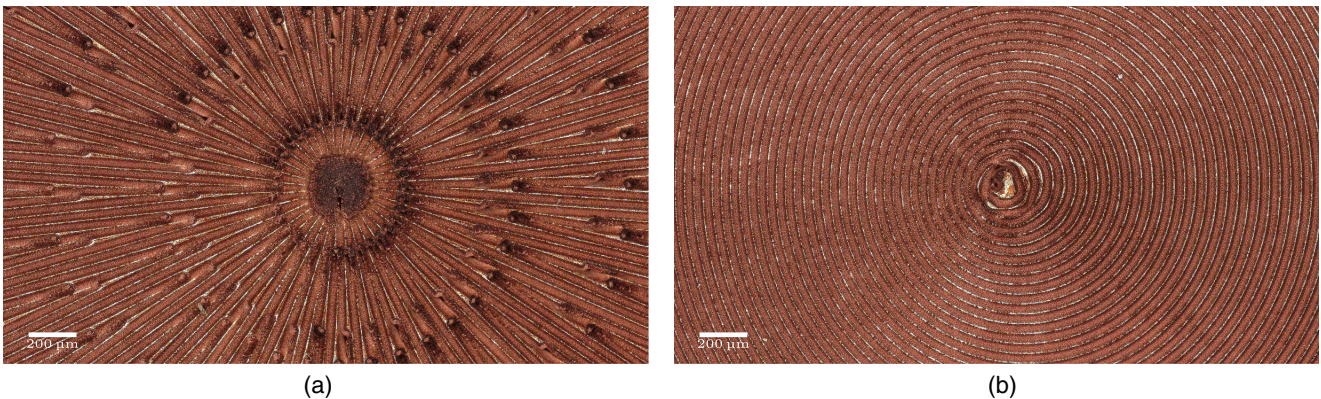

FIG. 1. Optical micrographs of the center of the laser-engineered surface structured copper discs. (a) Radial LESS-I structure and (b) Azimuthal LESS-II structure.

TABLE I. Processing stages and cleaning procedures used for modification of the surface with the intention of particle removal.

| Processing step | Cleaning procedure [all processes at room temperature (RT) if not otherwise indicated] |
|---|---|
| 0 | As received after laser processing |
| 1 | Blowing with 5 bar $N_2$ followed by 15 min rinsing with ultra-pure water and subsequent rinsing with ultrapure alcohol followed by $N_2$ dry blowing |
| 2 | Ultrasonication in deionized water at 150 W for 10 min followed by spraying with ethanol and dry blowing with filtered compressed air |
| 3 | Degreasing by immersion around 50 °C for approximately 2 h with ultrasonic agitation time of 15 min in detergent DP 17.40 SUP (NGL Cleaning Technology SA, Switzerland), followed by rinsing with demineralized water jet and by immersion, spraying with ethanol and dry blowing with compressed nitrogen and bake by hot air flow at 60 °C for up to 60 min, cooldown to RT (this corresponds to the typical surface treatment of copper components for use in ultrahigh vacuum at CERN) |
| 4 | All steps of cleaning process 3 followed by pickling with hydrochloric acid (50% v/v) for 10–20 s, rinsing with demineralized water jet and by immersion, passivation in aqueous solution of chromic acid (70–80 g/l) and sulfuric acid (3 ml/l) for 10–20 s (etching rate of about 1 μm/min), rising with demineralized water jet and by immersion, spraying with ethanol and dry blowing with compressed nitrogen and mild bake by hot air flow at 60 °C for up to 60 min, cooldown to RT |

### B. Cleaning procedures

The samples underwent an iterative examination before and after each consecutive cleaning step to verify the cleaning procedures' efficacy. This iterative approach involves subjecting the samples to repeated inspection and analysis following each cleaning stage, including optical microscopy, SEM analysis, XPS analysis, surface resistance determination, and SEY measurement. Through this approach, the progress of the cleaning process was evaluated, the quantity of remaining nanoparticles was determined, and changes in surface morphology were observed. This allowed to derive conclusions on the changes in physical surface properties.

Four different cleaning steps were performed to gradually remove nanoparticles and maximize the exposure of the groove pattern. The cleaning procedures for each step are detailed in Table I, which followed an approach of gradually increased particulate removal and cleaning capability by stepwise adding mechanical forces via the introduction of ultrasonication and later utilizing solutions which in addition chemically attack the surface (DP 17.40 SUP—low Cu etching rate of ∼20 nm/h, chromic acid/sulfuric acid mixture—Cu etching rate of ∼1 μm/min).

The cleaning step value (0–4) denotes the cleaning sequence and will be utilized for naming the samples. For instance, the samples obtained directly after the laser treatment are denoted as LESS-I-0 and LESS-II-0.

## III. CHARACTERIZATION OF SAMPLES

### A. Optical inspection

One of the key advantages of optical imaging is its ability to provide nondestructive and high-resolution imaging of samples, as shown in Fig. 2 for the LESS-II-0 sample. It allows for the visualization of both macroscopic and microscopic details for a quick inspection of the samples. Depicted in the figure are intricate details of the spiral pattern with clearly discernible redeposited particulates. In addition, the optical imaging enables the examination of the groove depths through three-dimensional reconstruction of the grooves, exemplified by the LESS-I-0 sample in Fig. 2. The height of the reconstructed profile corresponds to the cross-sectional perspective, as indicated by the turquoise line in the figures, from left to right. The determination of the groove depths has revealed an average relative depth per groove ranging from approximately 15 to 30 μm for LESS-I and 15 to 25 μm for LESS-II, depending on the location on the disc. Additionally, the anticipated laser treatment line distancing of 45 μm, previously mentioned in Sec. II A, has been verified.

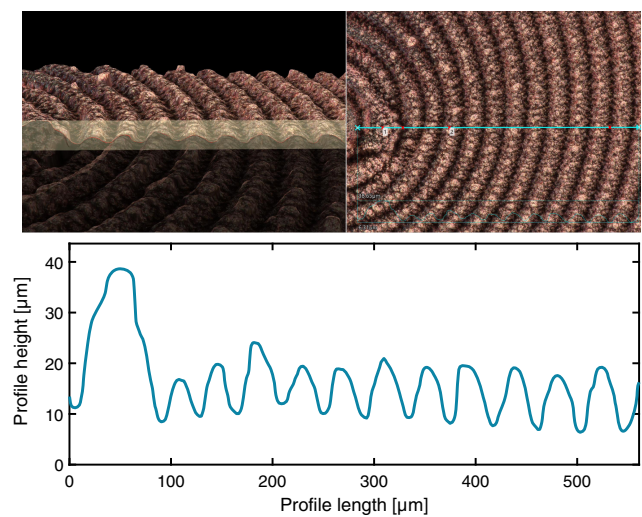

FIG. 2. Three-dimensional reconstruction and profile measurements of the LESS-II-0 configuration via optical imaging.

### B. SEM analysis

A more detailed view of the microstructure, compared to the optical inspection, can be achieved by SEM imaging. Figure 3 compares the microscopic analysis of the disc surfaces (approximately at the same locations, taken after the different processing steps) using SEM. Figure 3(a) presents the SEM results for LESS-I, while Fig. 3(b) shows the results for LESS-II, and Fig. 3(c) offers a more detailed examination of the microstructure by zooming in onto one of the grooves, in this case, for LESS-II. Each value in these three subfigures corresponds to the cleaning process given in Table I.

The findings suggest that the laser-treated samples directly after the LESS treatment contain a vast amount of agglomerated particles of different sizes up to 50 μm independent of the laser treatment direction. The first cleaning removed the aforementioned particle agglomerations, whereas the residual nanoparticles are more firmly bonded to the surface [16], and could not be removed this way. Due to the strong bonding, the density of nanoparticles attached to the surface after the second cleaning remains similar to before the second cleaning. The third cleaning step exposed the crests, and the efficiency of removing particles from the side walls and bottom of the grooves remained low. The images reveal that the fourth cleaning method is the most aggressive among the procedures employed. In this case, the copper microgroove structure is exposed, and the nanoparticles layer is almost completely dissolved, except for the groove valley, where the surface remains rough.

### C. Surface composition: XPS analysis

XPS is a surface-sensitive analytical technique that enables the determination of the elemental composition of a material at its outermost surface region with an information depth of only a few nanometers (5–10 nm, depending on the material and experimental condition) [24]. The details of the utilized experimental setup are described elsewhere, e.g. [16]. After each cleaning step, the samples were transported through the air and then introduced to the ultrahigh vacuum system for surface analyses via a load lock. Table II includes the quantity in at. % of detected elements. These values were calculated assuming a homogeneous lateral and horizontal distribution of all elements utilizing the tabulated photoelectron cross-sections of Scofield [25], considering an inelastic mean free path (IMFP) of the emitted electrons that is dependent on their kinetic energy: IMFP $\propto E_{\text{kin}}^{0.7414}$ [26], as well as considering the energy-dependent transmission function of the electron analyzer. Even though this simple calculation overestimates the quantity of species that are closer to the surface within the depth of information, specially surface adsorbates, the values allow to draw important conclusions on variations of the surface after each cleaning step, which is essential for interpreting the results of SEY measurements in the following section. One has to note that for the first processing (step 0) as well as for all the other cleaning steps, no noticeable difference was found between the two patterns inscribed. Therefore, only selected spectra of the radial pattern are shown in Fig. 4.

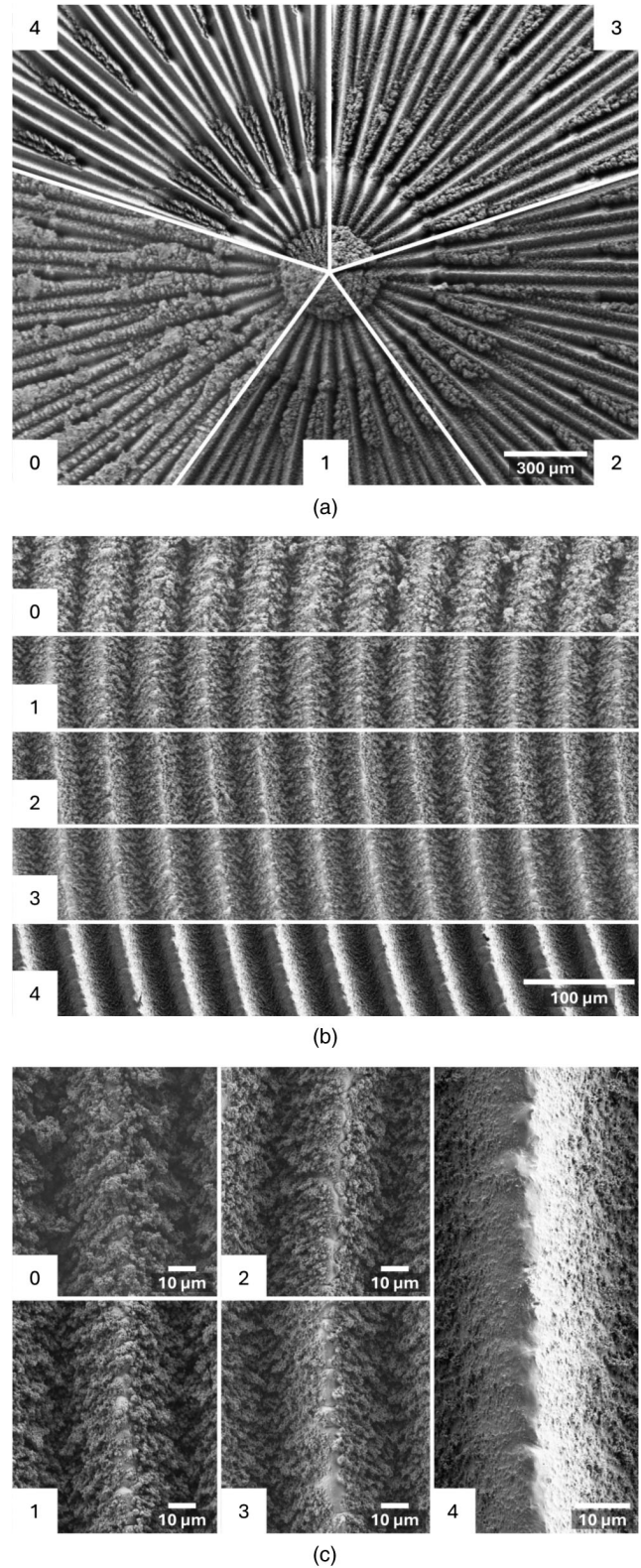

FIG. 3. SEM images of gradually cleaned laser-treated copper surface. Cleaning procedures are given in Table I. (a) Radial LESS-I pattern. (b) Azimuthal LESS-II pattern. (c) Detailed view on the grooves of the azimuthal pattern.

TABLE II. The surface composition of the detected elements in atomic percent (at. %) based on XPS measurements and quantitative analysis, which assumes a homogeneous distribution of all detected elements.

| Sample | Ag | C | Cr | Cu | N | O | S | Si |
|---|---|---|---|---|---|---|---|---|
| LESS-I-0 | | 20.7 | | 38.2 | 1.1 | 40.0 | | |
| LESS-II-0 | | 20.2 | | 41.3 | 3.2 | 35.4 | | |
| LESS-I-1 | | 17.4 | | 44.8 | 3.8 | 34.0 | | |
| LESS-II-1 | | 20.2 | | 42.3 | 2.8 | 34.7 | | |
| LESS-I-2 | | 35.4 | | 33.3 | 2.6 | 28.8 | | |
| LESS-II-2 | | 33.9 | | 33.8 | 2.7 | 29.6 | | |
| LESS-I-3 | 1.0 | 11.9 | | 46.5 | 2.7 | 34.1 | | 3.9 |
| LESS-II-3 | 0.6 | 14.5 | | 44.9 | 4.3 | 31.1 | | 4.6 |
| LESS-I-4 | 0.03 | 27.7 | 3.6 | 24.7 | | 39.4 | 1.3 | 3.1 |
| LESS-II-4 | | 32.4 | 3.9 | 22.7 | | 37.5 | 1.6 | 2.0 |

The laser-treated surfaces possess a composition that is influenced by surface reactions during the processing, which was performed in nitrogen, and by adsorbates due to exposure to air. As a consequence, the surface after processing exhibits a strong degree of oxidation (presence of $Cu_2O$ at the surface, the properties are in accordance with comparable samples discussed earlier [18,27] and references therein) as well as hydrocarbon species, while the nitrogen originates from reactions during the laser processing [28]. In addition, the spectrum includes signatures of surface hydroxide bonds [29], which are also formed by reactions in air. After the first rinsing step, which removed only the particulates that were loosely attached to the surface, no significant change was observed, while the second cleaning step led to a slight increment of the surface carbon content. This could be linked to the final rinsing in ethanol during this procedure. After the third cleaning step with a detergent that also slightly etches the surface, distinct changes are observed. The amount of surface hydroxide bonds is significantly reduced, the surface hydrocarbon content is much lower, and the Cu LMM Auger spectrum indicates the partial presence of metal Cu surface atoms (oxidation state $Cu^0$) [29,30]. Furthermore, the detected surface impurities of low content (Ag, Si) are linked to ingredients in the detergent solution, which typically cannot be completely rinsed off by water and, therefore, remain at the surface. The final cleaning step includes strong etching of the surface with hydrochloric acid and a passivation step including chromic acid, leading to the formation of chromic oxide at the surface. Figure 4 includes the signature of the chromium at the surface next to the Cu LMM transition, which also indicates an even higher relative content of metallic Cu atoms. On the other hand, the Cu 2p spectrum reveals that a partial buildup of hydroxide bonds took place in parallel. In addition, the final strong etching, intended to remove the nanostructure in the groove pattern, led to an increase in the hydrocarbon content and a removal of the Ag surface impurities, while the Si content remained almost constant. Finally, a slight amount of sulfur is detected, which is due to the utilization of sulfuric acid during this cleaning procedure.

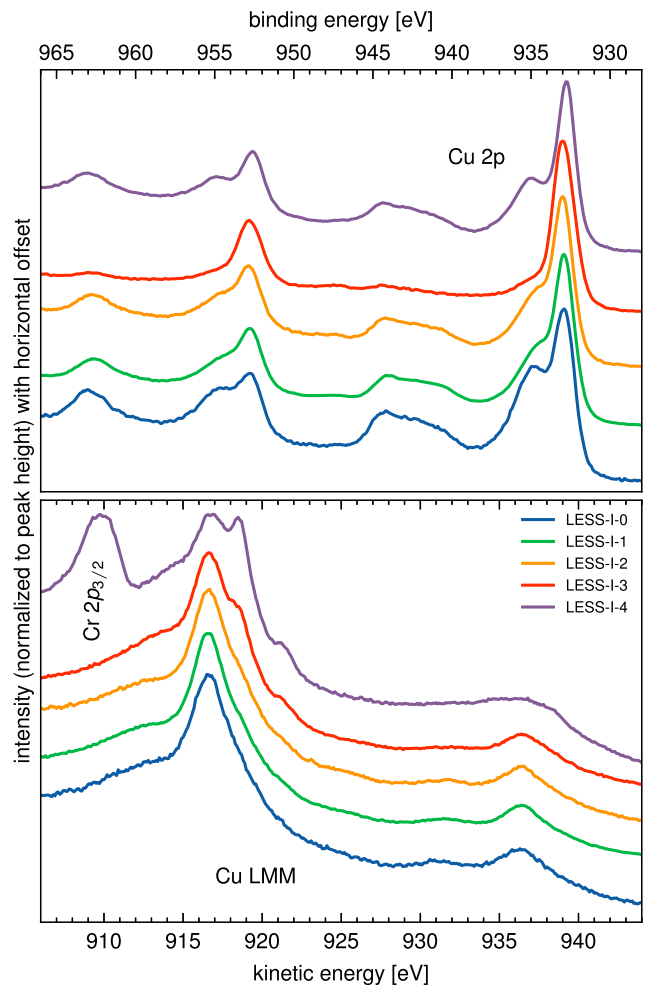

FIG. 4. Electron spectra of the Cu 2p (top) and Cu LMM (bottom) emission of the sample with radial pattern (LESS-I) after the different processing steps (indicated in the legend). The measurements of the sample with the azimuthal pattern (LESS-II) are similar and therefore not shown.

### D. Secondary electron yield

The secondary electron yield for primary electron energies between 50 and 1800 eV was measured on the samples by the method of alternating bias with experimental conditions as described earlier [18,31] prior to XPS analysis to avoid any x-ray induced conditioning effect. Figure 5 includes the results of the measurements as an average of seven spots for each sample and cleaning step, including the lateral SEY variation represented by the shown statistical uncertainties, as well as information on the SEY maximum $\delta_{max}$ and its energy $E_{max}$ as inset. Since the samples lacked circumferential markings, the positioning of the spots varied for each cleaning cycle due to possible sample rotation. The relatively large scattering of data indicates lateral inhomogeneity.

In general, the secondary yield can be influenced by two main aspects: (i) changes in surface roughness and topography and (ii) changes in the surface composition. Therefore, the results of SEM and XPS analyses are considered for the interpretation of the SEY measurements. During the first two cleaning steps, the slight removal of weakly attached nanoparticles leads to only a minor increment of the SEY curve $\delta(E)$ and minor changes in the elemental surface composition, except for a hydrocarbon uptake after cleaning step 2 and its removal after cleaning 3. For the third cleaning step, the SEY average values were slightly reduced again, while the major topography change of removal of nanoparticles on the top edge of the microgrooves would be expected to induce the opposite, even though the areal portion is low. Overall, the observed minor structural modifications do not significantly change the surface roughness that the incoming electrons experience. They impinge on a fissured, nanostructured surface, which has many channels where the electrons can be trapped and dissipate their energy by forward scattering [16,32]. The lateral variations of the SEY across the samples are larger than the changes induced by the consecutive processing steps 0 to 3 and, hence, clear correlations cannot be extracted from the experimental data. The geometrical and compositional changes seem not to give a conclusive effect on the SEY.

However, after the final strong surface etching procedure (cleaning step 4), which led to the complete removal of the nanoparticle layer on top of the microgrooves, the SEY dramatically increased by almost a factor of 2. The chemical transformation, i.e., the formation of a surface that is passivated by chromic oxide, cannot explain the increment since the SEY maximum of $CrO_2$ films is between 1.6 and 2.0 [33,34] and typically no difference in SEY is found for flat Cu samples that have been treated with cleaning procedure 3 or 4, although their significant differences in surface composition. Consequently, this result indicates that the removal of the nanoparticles from the laser-formed microgroove structure induces a strong increase in secondary electron yield. This observation illustrates the important contribution of the redeposited nanoparticles on the SEY reduction effect generated by the laser processing.

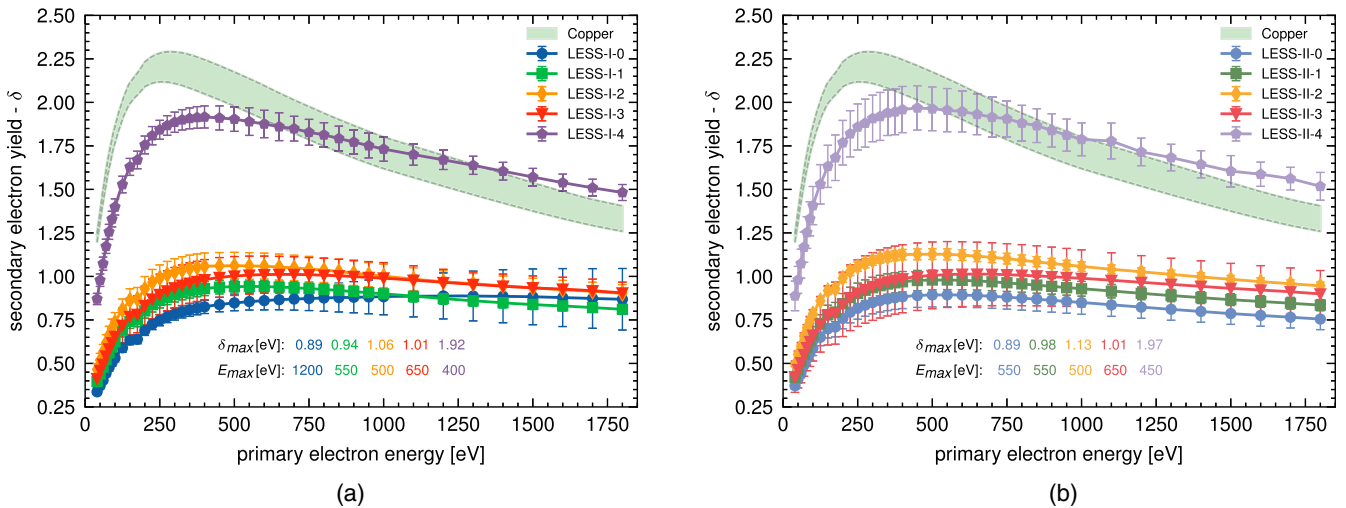

FIG. 5. Secondary electron yield in the range between 50 and 1800 eV for the radially and azimuthally laser-patterned samples after each surface cleaning step. The error bars represent the statistical measurement uncertainties for a confidence level of 68.3% based on seven measurements performed in different locations of the samples. The typical variation of the SEY of degreased (cleaning procedure 3) flat copper is included as a green region for comparison. (a) Radial LESS structure and (b) azimuthal LESS structure.

It should be noted in this context that the efficiency of particle removal depends on the structures that are created during laser structuring, the amount of readsorbed particles, and how well they adhere to the underlying surface. Their surface density can vary significantly when changing the processing conditions [17]. In the presented case, the initial surface is completely covered by a dense layer of well-attached particles, which could not be easily removed by rinsing or by treatment in an ultrasound bath. Especially the results of mechanical cleaning approaches such as ultra-sonication can vary when changing the applied power or cleaning time [35]. Thus, the reported results of our study on how efficient the particles are removed in step 2 have to be related to the applied cleaning conditions as described above. These aspects require consideration when comparing reported SEY values of laser-structured surfaces and the efficiency of particle removal and its consequences on SEY in different studies.

### E. Surface resistance assessment

The method selected for measuring the surface resistance $R_S$ is based on the closed Hakki-Coleman dielectric resonator (CHCDR) [36]. The design, as depicted in Fig 6 and explained in detail in [37], consists of a cylindrical copper cavity with dimensions of 105 mm in diameter and 20 mm height and is loaded with a low-loss c-oriented cylindric sapphire single crystal. The crystal's dimensions measure 40 mm in diameter and 19.5 mm in height. The test sample axially shields the cavity, with the processed surface of interest facing the crystal. To prevent any imprint of the single crystal onto the laser-treated samples, its height is reduced by 0.5 mm compared to that of the cavity. The resonator operates in the $TE_{011}$ mode with a resonance frequency of $f_0 = 3.4$ GHz.

The $R_S$ measurements were conducted across a temperature range from close to room temperature down to almost liquid nitrogen temperature (77 K). Since the electrical resistivity of copper decreases with lower temperatures, the rf skin depth (indicating the depth at which an electromagnetic wave penetrates a material before its intensity is reduced by $1/e$) decreases in the given temperature range from 1.2 to 0.5 μm. A stable resonance frequency is maintained due to the presence of the sapphire dielectric, which changes the relative permittivity between $\varepsilon_r = 9.4$ room temperature (RT)] and 9.1 (77 K), equivalent to a change of resonance frequency of $\Delta f_0 \approx 20$ MHz. It is important to note that although changes in permittivity are not the only factors affecting resonance frequency [38], they are the most significant in this particular setup. However, this characteristic facilitates the examination, particularly regarding the influence of nanoparticles on the surface resistance. The skin depth is primarily influenced by the measurement frequency and the material's conductivity. The stable resonance frequency creates a controlled environment that accommodates changes in conductivity due to temperature variations without interference from other variables, such as frequency shifts.

Figure 7 shows a normalized density plot of the surface magnetic field magnitude distribution on the sample, equivalent to the surface current density. As can be seen, the spatial variation of the field distribution of the $TE_{011}$ mode is mainly focused in the center of the sample close to the dielectric edge, though with a minimum in the sample center. The maximum magnitude of the purely azimuthal current density is located in the radial position of approximately 13 mm from the center. Consequently, for the LESS-I disc, the lines etched by the laser beam are orthogonal to the rf currents in this configuration. In contrast, the groove lining and the rf surface currents are aligned and almost parallel for the LESS-II disc. The impact of the nanoparticles on the impedance is generally believed to remain independent of the groove orientation and introduces an additional resistance in both line directions.

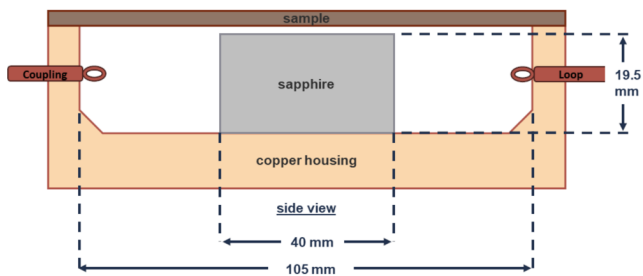

FIG. 6. Schematic cross-sectional view of the CHCDR. It consists of a copper cylinder that encases a sapphire dielectric at its center. The sample closes the cavity. Electromagnetic coupling is achieved through the lateral walls.

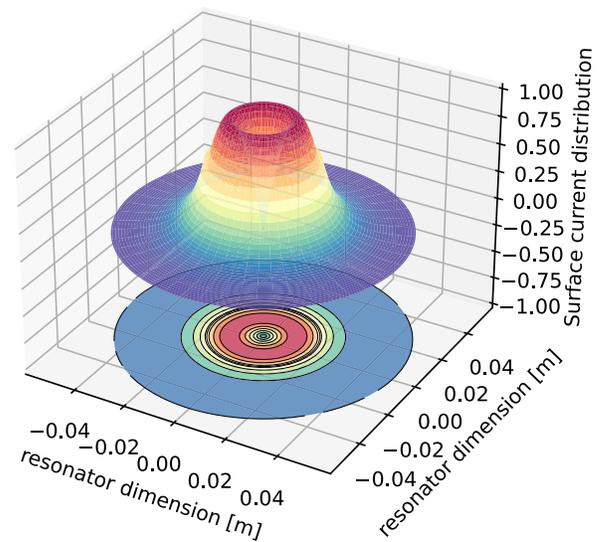

FIG. 7. Normalized rf-surface magnetic field distribution density plot ($TE_{011}$-mode) at the samples. The field strength is normalized to the maximum value. The red circle indicates the edge of the dielectric.

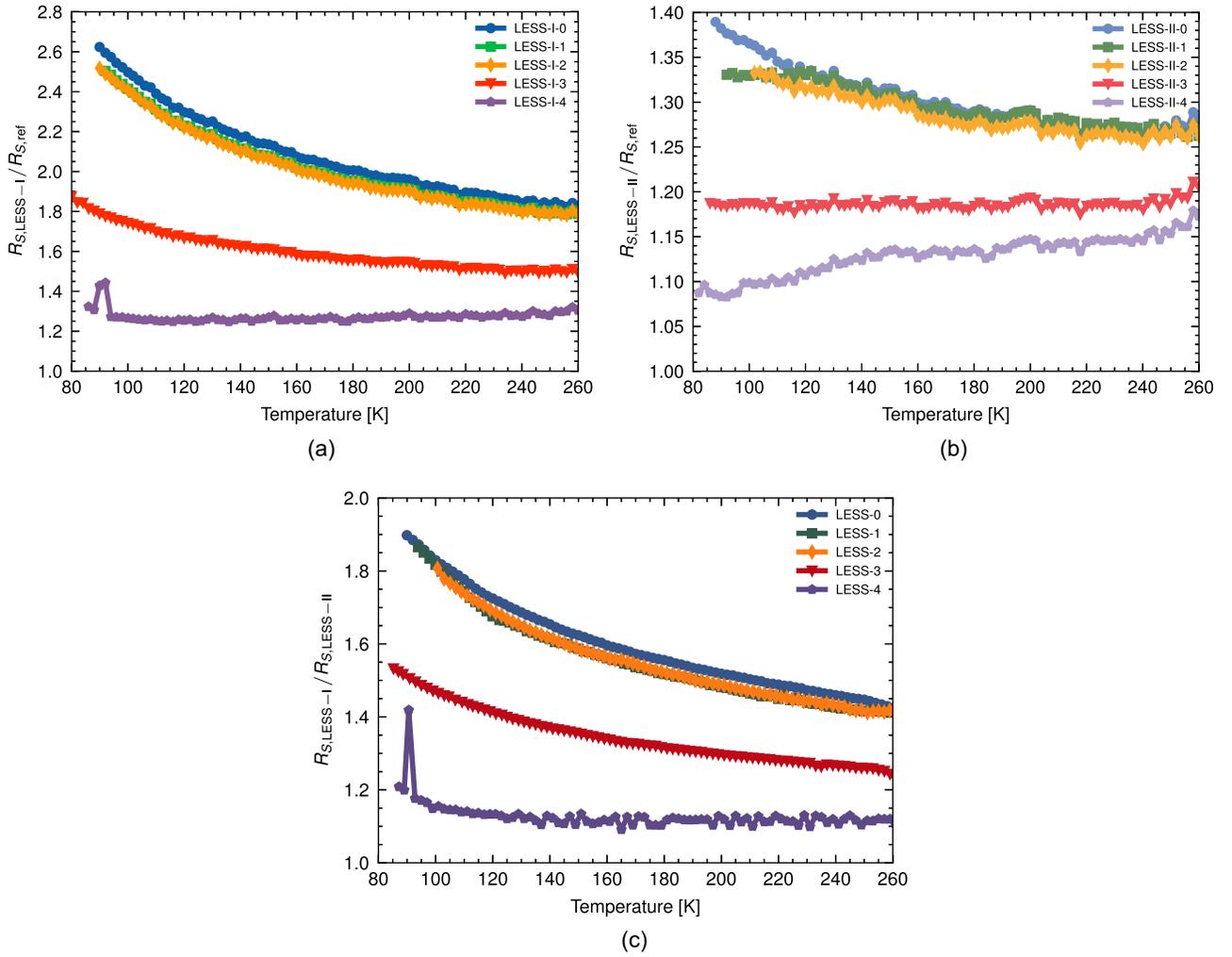

FIG. 8. Surface resistance ratios as a function of temperature for (a) radial LESS-I structure, (b) azimuthal LESS-II structure, and (c) direct comparison between the LESS structures for all cleaning steps at $f_0 = 3.4$ GHz.

The results in Fig. 8 illustrate the measured ratio in surface resistance $R_S/R_{S,\text{ref}}$ as a function of temperature for (a) the LESS-I sample and (b) the LESS-II sample, where $R_S$ refers to the measured surface resistance of the LESS samples and $R_{S,\text{ref}}$ refers to the surface resistance of a degreased flat OFE copper bulk disc (typical roughness $Ra \leq 3.2$ μm) used as reference, with examined $R_{S,\text{ref}}(\text{RT}) = 15$ mΩ and $R_{S,\text{ref}}(77\ \text{K}) = 6.7$ mΩ. For comparison, using the analytical expression to determine the surface resistance of a semi-infinite planar conductor with a smooth surface, $R_{S,\text{theo}} = \sqrt{\pi\mu_0\rho f_0}$, the theoretical room temperature value can be calculated as $R_{S,\text{theo}}(\text{RT}) \approx 15$ mΩ, assuming a copper resistivity of $\rho = 1.7$ μΩ cm [39]. At cryogenic temperature, the surface resistance is expected to decrease due to the reduction in resistivity. By applying the same theoretical framework and assuming a residual resistivity ratio (RRR) between $\text{RRR} = 10$ and $\text{RRR} = 100$, i.e. $\rho_{\text{RRR}=10}(77\ \text{K}) = 0.4$ μΩ cm and $\rho_{\text{RRR}=100}(77\ \text{K}) = 0.2$ μΩ cm [40], the theoretically expected surface resistance ranges from approximately $R_{S,\text{theo}}(77\ \text{K}) \approx 7.3$ to $5.2$ mΩ, respectively.

As can be seen in Fig. 8, each set of results comprises five measurements following the cleaning procedures outlined in Sec. II B. Two main observations emerge from the data: first, the surface resistance increases as the angle between surface currents and groove direction grows, and second, $R_S$ decreases when the laser-treated surface is cleansed of nanoparticles. As expected, nanoparticles, like surface roughness, increase surface resistance [41]. Both sets exhibit similar trends overall. A cleaner surface with a lower nanoparticle density corresponds to reduced overall roughness. Removing agglomerated particles consistently leads to a minor reduction, regardless of the groove orientation. Furthermore, ultrasonic cleaning shows no significant improvement, in agreement with the microscopic examination in Fig. 3. Notably, the relative increase of $R_S$ compared to degreased OFE copper ranges from approximately a factor of 1.1 to 1.4 for the azimuthal-treated sample and from 1.2 to 2.8 for the radial-treated sample. The efficacy of cleaning steps 3 and 4 in inducing changes in $R_S$ surpasses that of the other two cleaning steps by a significant margin. The LESS-II-4 sample attains even a level of performance comparable to that of the OFE copper reference sample.

Figure 8(c) shows the direct comparison of the surface resistance of the two laser-treated samples $R_{S,\text{LESS-I}}/R_{S,\text{LESS-II}}$. The results indicate that the ratio on the one hand increases as the temperature decreases. On the other hand, the ratio decreases with each cleaning step. These findings suggest that the influence of the nanoparticulates on the surface impedance cannot be seen as an additional constant term in combination with the geometrical orientation of the microgrooves.

## IV. DISCUSSION

The observed dynamic effects in the results of SEY and surface resistance support to a certain extend the general contradictory performance of these two parameters; low surface resistance is associated with high SEY values and vice versa. Nonetheless, the findings of this study suggest as well that the surface resistance is more sensitive to partial particle removal and that depending on the cleaning procedure, SEY and surface resistance values can be modified to meet the specific requirements of the intended application environment.

The measurement results imply that the macrogeometry that is generated by the laser treatment, precisely the azimuthal and longitudinal groove orientation, significantly impacts $R_S$ but not the SEY value. The uncleaned LESS samples exhibited significant agglomeration of particles, resulting in the highest relative surface resistance compared to OFE copper. The impact of the disparity between the rf current direction in parallel and perpendicular orientations is evident in the surface resistance values, showing differences up to a factor of 2. This ratio is half of the one that was measured in a quadrupole resonator [19] at lower temperatures of 4.2 K. Considering the data in Fig. 8(c) and the increment of the resistance ratio at lower temperature could likely lead to comparable values when extrapolating the data further to < 10 K. Furthermore, even though the laser pulse length, wavelength, and repetition rate during the processing of the different samples are the same in both studies, other laser parameters such as scanning speed, spot diameter, and groove spacing are different. As a consequence, the structural parameters—height, width, inclination angle, and aspect ratio—of the formed microgrooves differ and likely alter the surface resistance. When including frequency dependency and the combined effects of grooves and nanoparticles, theoretical modeling can help to understand the differences and improve comparability [42]. In contrast, the SEY value directly after processing is lower than that after all subsequent cleaning steps and irrespective of the laser structure pattern. Clearly, the groove orientation does not change the SEY for perpendicularly impinging primary electrons.

Microgeometry factors, including particulate sizes and the areal coverage of particulates, play a substantial role in influencing both parameters. The initial cleaning step eliminated the agglomerations and larger loosely attached particulates on the samples. The removal of these nanoparticles resulted in negligible changes in the elemental quantities of the surface topography. As a consequence, a minor increase in the SEY occurred while the surface resistance experienced a slight reduction (regardless of the groove orientation).

The third cleaning step primarily exposed the crests, while the grooves underwent insignificant alteration. Reduction of particulates distribution on the crests, i.e., a local decrease of surface roughness, would provide more opportunities for primary electrons to interact with the underlying Cu substrate. However, the lateral area that is affected is rather small, while the roughness of the remaining regions is unchanged. The found variation in the SEY is within the lateral variation across the sample and no clear trend can be extracted. On the other hand, the drastic reduction of $R_S$ implies that the surface current predominantly flows along the crests, with a potential gradual surface current density distribution over the height of the grooves. Removing the particulates from the crests and exposing the bulk copper could account for the observed reduction in surface resistance. These observations suggest that the particulates are of good conducting material (copper), and their dimensions are larger than the penetration depth, as observed in the SEM images of Fig. 3, thus impacting the surface resistance. Contrarily, the SEY value is significantly driven by the remaining nanoparticles inside the grooves. The optimal scenario appears to be having clean crests for low surface resistance, coupled with a gradual increase in nanoparticles within the grooves to maintain a low SEY value, and is feasible using the detergent cleaning including mild etching. This observation could undergo cross validation in the future through measurements conducted at various frequencies to ascertain a potential correlation between the exposure of groove tips and penetration depths.

Besides, the influence of the two relevant interaction depths shall be taken into consideration. First, the main region from which secondary electrons are emitted from is several tens of nanometers from the surface. This aspect makes the electron emission relatively sensitive to surface composition changes for flat materials such as several nanometer thick oxide layers. Second, the electromagnetic penetration depth ranges in the micrometer scale. Thus, for the surface resistance measurements, such oxide layers with low electrical conductivity are "invisible" to the electromagnetic field since the skin depth is much larger than the thickness of the oxide layer. Thus, natural oxide layers do not influence the measured surface resistance. Nevertheless, in both cases, SEY and $R_S$, the additional aspect of surface roughness caused by the nanoparticles makes the electron scattering and electromagnetic interaction processes more complicated and less theoretically predictable.

The most aggressive cleaning method, the fourth step involving strong etching and passivation, induced a transformative topography change. Complete removal of nanoparticles from the grooves and exposure of the underlying copper microgroove pattern resulted in a dramatic increase in SEY, almost doubling its value. This finding reassures that the nanoparticles on the microstructures, primarily at the grooves, are the essential ingredient for the capability of LESS to reduce the SEY, while the triangular cross section of the remaining microgrooves does not exhibit a very different secondary electron emission characteristic than a flat Cu sample of identical surface composition. It has to be noted that this observation is only validated here for the structural dimensions characterized, while other microscopic geometries, i.e., especially steep structures with higher aspect ratio, are modeled and demonstrated to also reduce the secondary electron emission [43–45].

## V. CONCLUSION

The presented study provides insights into the intricate interplay between surface morphology; nanoparticle deposition; cleaning procedures; and resulting physical, chemical, and electromagnetic properties of laser-treated copper surfaces. The insights gained from this investigation may be of assistance to optimize laser surface structuring techniques for various applications and offer potential avenues for further research in this field. The distinct impacts of macrogeometries (surface morphology and groove depth and orientation) and microgeometries (nanoparticle sizes and area coverage) are revealed, focusing on the implications for secondary electron yield and surface resistance dynamics. The results indicate that both parameters can be modified by cleaning approaches depending on the application's specific requirements despite the typical contradictory performances of these two parameters. The significant reduction in surface resistance at constant SEY values suggests that a combination of exposed grove crests while retaining particulate coverage in the grooves creates an effective balance between these two parameters. The exposed groove crests cause a significantly reduced surface resistance compared to the case where particulates also cover the crests. This observation suggests a gradual surface current density distribution along the groove height, potentially influenced by other factors such as measurement frequency and related variation of the skin depth. This observation, nevertheless, warrants further theoretical and experimental investigation and validation in this area. The aggressive cleaning method, involving strong etching and passivation, induced a transformative change by completely removing the nanoparticles and exposing the underlying copper substrate. The remaining structure of the microgrooves exhibited secondary electron emission characteristics that are similar to those of a flat copper sample. This emphasizes the crucial role of nanoparticles in reducing the SEY of laser-treated surfaces.

## ACKNOWLEDGMENTS

S. W. and A. A. at the University of Dundee would like to express their gratitude to the Science and Technology Facilities Council (STFC) of the United Kingdom for support through Grant No. ST/T001887/1. Authors from CERN express gratitude to K. Brunner for the development and implementation of the cryogenic dielectric resonator system, which involved surface resistance measurements on laser-treated samples during preparatory stages.